\documentclass[aps,prl,twocolumn,showpacs]{revtex4}
\usepackage{amsmath,amsthm}  
\usepackage{epsf}
\usepackage{amssymb}  
\usepackage{graphicx}

\newcommand{\be}{\begin{eqnarray}}
\newcommand{\ee}{\end{eqnarray}}
\newcommand{\icum}[1]{\langle\langle I^{#1}\rangle\rangle}
\newcommand{\eq}[1]{Eq.~(\ref{#1})}
\begin{document}
\title{Feedback Control of Quantum Transport}
\author{Tobias Brandes}
\affiliation{  Institut f\"ur Theoretische Physik,
  Hardenbergstr. 36,
  TU Berlin,
  D-10623 Berlin,
  Germany}
\date{\today{ }}
\begin{abstract}
The current  through nanostructures like quantum dots can be stabilized by a feedback loop that continuously adjusts system parameters as a function of the number of tunnelled particles $n$. At large times, the feedback loop freezes the fluctuations of $n$ which leads to highly accurate, continuous single particle transfers. For the  simplest case of  feedback acting simultaneously on all system parameters, we show how to  reconstruct the original  full counting statistics from the frozen distribution. 
\end{abstract}
\maketitle

Fluctuations of the electronic current   have become a major tool for probing quantum coherence, interactions, and  dissipation effects in quantum transport through nanoscale structures \cite{Naz03}. Monitoring quantum objects  during their time evolution 
usually introduces extra noise, but it can also compensate backaction effects and be used for recycling information 
in order to control the system dynamics \cite{Wiseman_Milburn}.

The random tunnelling of electrons in quantum transport is described by the full counting statistics (FCS) of transferred charges.  Similar to equilibrium thermodynamics where, e.g., the cumulants of the particle number distribution in the grand canonical ensemble are proportional to the volume,  FCS cumulants in stationary transport linearly increase in time (exceptions are possible near phase transitions \cite{KvO10}). All quantum transport devices thus have to deal with a stochastic element that can become a major obstacle when very regular currents are required. 

Here, we show that this situation changes by  `freezing' the cumulants in time, if one applies feedback (closed loop) control \cite{Wiseman_Milburn} to  quantum transport. We propose a scheme where a time-dependent signal $q_n(t)$ is used to continuously adjust system parameters such as tunnel rates or energy levels. 
Here,  $q_n(t) \equiv I_{0}t  -n$ is  an error charge determined from the ideal `target' current $I_{0}$ and the total charge $n$ that has been collected in  (or flown out of) a reservoir during the measurement (e.g., by a nearby quantum point contact detector) up to time $t$. The error charge determines whether to speed up or slow down the transport process -- a  form of feedback that is analogous to the centrifugal governor used, e.g., in thermo-mechanic machines like the steam engine. 

We describe transport in the usual way by coupling the system Hamiltonian $\mathcal{H}_S$ of a few-state nanostructure (e.g., a quantum dot) to left and right reservoirs $\mathcal{H}_{L/R}$  via a tunnel Hamiltonian $\mathcal{H}_T$. The feedback loop is modelled by an dependence of the parameters in $\mathcal{H}_S$ and $\mathcal{H}_T$ on time $t$ and on the number operator $\hat{N}_R$ of the right reservoir (drain). In the usual Born-Markov approximation in lowest order in  $\mathcal{H}_T$, the reduced system density operator $\rho^{(n)}(t)$ conditioned on the number of electrons $n$ tunneled from left to right  (we assume a high-bias situation with unidirectional transport) obeys a Master equation
\be\label{master_tunnel}
\dot{\rho}^{(n)}(t) =\mathcal{L}^0_n(t) \rho^{(n)}(t) + \mathcal{J}_{n-1}(t) \rho^{(n-1)}(t).
\ee
Here, $\rho^{(n)}(t)$ is a vector with $d$ real components representing system occupations and coherences, and 
in contrast to the usual $n$-resolved Master equations \cite{GP96}, 
the jump ($\mathcal{J}$) and non-jump ($\mathcal{L}^0$) super-operators ($d\times d$ matrices) 
have a time- and $n$-dependence \cite{Milburn_footnote} which
can in principle be derived from a microscopic model for, e.g., the tunnel matrix elements $V_k= V_k(t,\hat{N}_R) $ in $\mathcal{H}_T$, or from a corresponding dependence of the energy levels of $\mathcal{H}_S$. 
In an experiment, one could use the signal of a quantum point contact in combination with an electronic circuit to modulate, e.g.,  the gate voltages that determine the tunnel rates $\Gamma_{L/R}(n,t)$ between the reservoirs and the nanostructure.

In all what follows we will assume the elements of $\mathcal{L}^0_n(t) $ and $\mathcal{J}_{n-1}(t)$ multiplied by analytic functions 
\be
f(q_n(t)),\quad q_n(t)\equiv I_{0} t - n,\quad f(0)=1
\ee
that describe the modulation of system parameters by the feedback loop, requiring non-invasive feedback with no modulation for zero error charge $q_n(t)=0$.
We also assume $I_{0}=I$ as the stationary current $I$ without feedback in order  to simplify some of the formulae; the case $ I_{0}\ne I$ yields analogous  results.

Let us start with the simplest transport model:  a tunnel junction with no internal system degrees of freedom ($d=1$) and  $-\mathcal{L}^0_n(t)=\mathcal{J}_n(t) \equiv  \Gamma \times f( q_n(t))$, where $\Gamma$ is the rate for tunnelling of electrons from left to right.  We first consider  {\em linear feedback} $f(x)=1+gx$, where $g$ is  a dimensionless feedback parameter. This form  is appropriate for weak feedback coupling $g\ll 1$. A simple calculation then yields the first two cumulants of the FCS $\rho^{(n)}(t)$ as the average $C_1(t) \equiv \langle n\rangle_t= \Gamma t$,  and the variance
\be\label{variance_tunnel}
C_2(t) \equiv \langle n^2\rangle_t- \langle n\rangle_t^2 =
\frac{1}{2g}\left( 1- e^{-2g\Gamma t}\right). 
\ee
This already shows that at any finite feedback strength $g>0$, there occcurs a drastic change:
the cumulants $C_k$, $k\ge 2$ no longer increase linearly in time $t$ but converge to a constant, e.g. $C_2(\infty) = \frac{1}{2g}$. This means that the FCS charge distribution no longer spreads out but freezes into a stationary distribution with a fixed shape that constantly moves to larger $n$, with a mean value  $\langle n \rangle_t = I_{0} t$, cf. Fig. {(\ref{figure1}a).

\begin{figure}[]
\includegraphics[width=\columnwidth]{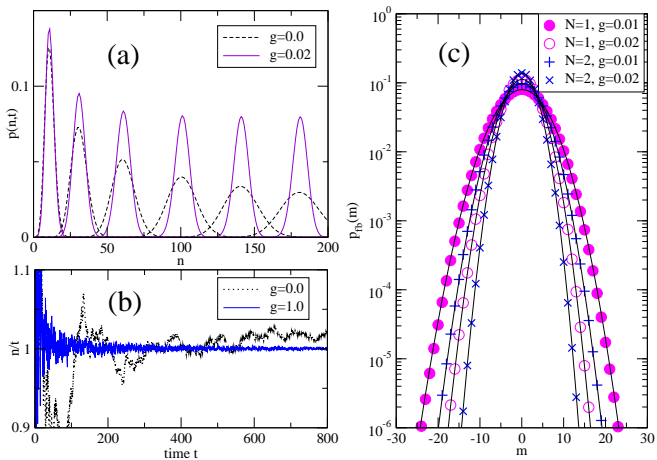}
\caption[]{\label{figure1}
a) Distribution of number $n$ of tunneled particles at times $t = 30, 60, 100, 140, 180$  for tunnel junction (tunnel rate $\Gamma=1$, linear feedback strength $g$) ; b) Single realisation (trajectory) of $n/t$ for tunnel junction as a function of time $t$ ($\Gamma=1$); c) frozen feedback distributions $p_{\rm fb}(m)\equiv \lim_{t\to \infty} p(\langle n \rangle_t +m  ,t)$  for chain with $N=1$ and $N=2$ quantum dots, symbols: numerical solution of \eq{master_tunnel}, lines from \eq{pfb}and re-scaled \eq{tunnel_analytics}, see text. 
}
\end{figure} 

In order to obtain $\rho^{(n)}(t) $ and all the other cumulants $C_k$, we define the Fourier transform $\rho(\chi,t)\equiv \sum_n \rho^{(n)}(t) e^{i\chi n}$  and use \eq{master_tunnel} to 
derive the partial differential equation 
\be\label{PDE}
\frac{\partial}{\partial t }{\rho}(\chi,t) = \mathcal{L}(\chi) f\left(I_{0}t-\frac{\partial}{\partial i\chi}\right){\rho}(\chi,t),
\ee
where $\mathcal{L}(\chi)\equiv  \Gamma (e^{i\chi}-1)$ and $I_{0}=\Gamma$ (we set the elementary charge $-e=1$). 
From the solution of \eq{PDE}, one finds the cumulant generating function (CGF)  $\mathcal{F}(\chi,t)$ for linear feedback,
\be\label{tunnel_analytics}
& &\mathcal{F}(\chi,t)= I_{0} t {i\chi} +\frac{1}{g}  \ln \left( e^{i\chi} (1-e^{-g\Gamma t})  +    e^{-g \Gamma t}     \right)
   \nonumber\\
&+&  \frac{1}{g}\left[ Li_2\left(\left(1-e^{-i\chi} \right)e^{-g \Gamma t}\right) - Li_2\left(1-e^{-i\chi}  \right)\right] ,
\ee
where $Li_2(z) \equiv\int_z^0 \frac{dt}{t}\ln(1-t)$. From \eq{tunnel_analytics}, we find explicit expressions for the FCS via $\rho^{(n)}(t) = \int_{-\pi}^{\pi}\frac{d\chi}{2\pi}e^{-in\chi}e^{\mathcal{F}(\chi,t)}$, and for the cumulants $C_k(t) \equiv \frac{\partial^k}{\partial(i\chi)^k}\mathcal{F}(\chi,t)|_{\chi=0}$. In the long time limit, one obtains
\be\label{cumulantstunnel}
C_k(t\to \infty) =  -\frac{1}{g}B_{k-1},\quad k \ge 2,
\ee
where the $B_k\equiv \frac{d^k}{dx^k}\left.  \frac{x}{e^x-1}\right|_{x=0}$ are the $k$-th Bernoulli-Seki numbers (the cumulants thus grow rapidly at large $k$ \cite{Flietal09}).
The presence of feedback thus transforms the originally Poissonian FCS ($C_k=\Gamma t$ for $g=0$) into a non-diffusive, constantly moving distribution at large times. This qualitative change is underlined by the fact that the finite-feedback results \eq{tunnel_analytics}, \eq{cumulantstunnel} are non-perturbative in the feedback coupling parameter $g>0$.  We found a further characteristic feature by solving the $n$-resolved Master equations \eq{master_tunnel} via the quantum jump method \cite{carmichael}, which in the tunnel junction ($d=1$) case amounts to a simple stochastic algorithm simulating individual experimental realisations of electron tunneling histories $n(t)$, cf. Fig. (\ref{figure1}b). On long time scales, feedback suppresses large deviations of $n(t)$ (we plot $n(t)/t$ for a clearer picture) that  without feedback lead to the linear increase of the cumulants $C_k(t)$ with time. In contrast, on short time scales this distinction is barely visible, which  is also underlined by the fact that the {\em waiting time} distribution of electron tunneling $w(\tau)$ \cite{carmichael} reacts much less sensitive to feedback (not shown here).

As a next step, we elucidate the role of internal system  degrees of freedom.
The simplest case here are spin-polarized electrons with sequential tunneling through a chain of $N\ge 1$ single resonant levels.  We assume strong Coulomb blockade, i.e. only one additional electron on the chain at a time, and  tunnel rates 
\be\label{rates}
\Gamma_\alpha(n,t)\equiv \Gamma_\alpha \times f(1+g_\alpha(I_{0}t-n))
\ee
that are simultaneously modulated. Results of the numerical solution of \eq{master_tunnel} are shown in Fig. (\ref{figure1}c) for identical 
rates $\Gamma_\alpha=\Gamma$. The FCS distributions 
\be \label{pfb}
p_{\rm fb}(m)\equiv \lim_{t\to \infty} \int_{-\pi}^{\pi}\frac{d\chi}{2\pi}e^{-i[ \langle n\rangle_t+m] \chi}e^{\mathcal{F}(\chi,t)}
\ee
are frozen around their maxima  that describe moving electron number averages $\langle n\rangle_t = I_{0}t$ in the large time limit. 

Again, we corroborated our numerical results by an analytical solution of the feedback master equation for the Fourier transformed density operator as in \eq{PDE}, where now  ${\rho}(\chi,t)$ is a $N+1$-dimensional vector and the Liouvillian matrix
$ \mathcal{L}(\chi)$ has entries $-\Gamma$ on the diagonal, $\Gamma$ on the lower sub-diagonal and $\Gamma e^{i\chi}$ in the right top corner, corresponding to counting electrons after they leave the $N$-th dot in the chain to the right reservoir.  At large times $t$, it turns out that $\mathcal{L}(\chi)$ can be replaced by its critical eigenvalue $\lambda_0(\chi)$ with $\lambda_0(0)=0$. This  approximation is in analogy to the long-time limit of the FCS for $g=0$ (no feedback), but for $g>0$  it strongly relies on the homogeneous coupling of the  feedback function to all matrix elements in $\mathcal{L}(\chi)$. 
Rescaling the tunnel rate as $\Gamma = \gamma (N+1)$, we find the CGF ${\mathcal{F}(\chi,t)}   \equiv  \ln {\rm Tr}_S \rho(\chi,t) $ as the r.h.s. of \eq{tunnel_analytics} with $I_{0}=\gamma$  and the replacements $g \to g/(N+1)$ and $\chi \to \chi/(N+1)$. The feedback-frozen distributions $p_{\rm fb}(m)$, \eq{pfb},
obtained in this way  are in excellent agreement with the numerical results. We mention that we carried out this analysis for the linear feedback case that is appropriate for $g\ll 1$. In general, modelling $f$ as an $n$-th order polynomial leads to $n$-th order PDE systems which for $n>1$ make an analytical treatment cumbersome.

The  appearance of the eigenvalue $\lambda_0(\chi)$ in the above analysis suggests that for transport through an arbitrary system $\mathcal{H}_S$, the original FCS $p(n,t)$ without feedback can be reconstructed from the feedback-frozen $p_{\rm fb}(m)$ at large times $t$. This can be verified by writing the CGF as ${\mathcal{F}(\chi,t\to \infty)}  = i\chi I_{0} t + h(\chi)$, leading to
\be\label{relation}
\frac{i\chi}{\lambda_0(\chi)} I_{0}=  e^{-h(\chi) } f \left( -\frac{\partial}{\partial i\chi}\right) e^{h(\chi) }.
\ee
This, however, is only valid for feedback functions $f$ multiplying the Liouvillian $\mathcal{L}(\chi)$  as a whole, cf.  
 \eq{PDE} (`homogeneous feedback'). 
From \eq{relation}, explicit relations between the $g=0$ current cumulants $\icum{k}\equiv \frac{\partial^k}{\partial (i\chi)^k} \lambda_0(\chi)|_{ \chi=0}$  and the stationary feedback cumulants $C_k\equiv \frac{\partial^k}{\partial (i\chi)^k} {\mathcal{F}(\chi,t\to \infty)} |_{ \chi=0}$ follow: for weak feedback (linear feedback function), the first cumulant 
\be\label{C1}
C_1 = I_0 t + \frac{1}{g}\left( 1- \frac{I_0}{\icum{1}} \right)
\ee
describes the mean  target value $I_0 t$  of the feedback FCS  at large times $t$ plus an extra charge that flows into the drain reservoir due to the mismatch between the target current $I_0$ and the no-feedback stationary  current $\icum{1}$. If the latter is known (e.g., by letting the system run without feedback for a certain time), \eq{C1} can be used to determine the value of the feedback parameter $g$ in an experiment. 

Second, from \eq{relation} one recovers  the $g=0$ Fano factor $F_2\equiv \icum{2}/\icum{1}$ from the frozen $g>0$ cumulants as 
\be\label{fano_predict}
 F_2 =  2g C_2
\ee
for linear feedback, plus small corrections like $O(g_2^2)$ for feedback functions $f$ with  quadratic ($g_2x^2$) or higher terms  in their Taylor expansion around $x=0$.   
Corresponding relations  can be easily derived from \eq{relation} for higher cumulants, e.g., the skewness for linear feedback $ F_3 \equiv \icum{3}/\icum{1}= 6g^2 C_2^2 + 3g C_3$.

We tested these results and extended them to non-homogeneous feedback by considering the single resonant level model, where at large bias voltage the Master equation reproduces the exact solution for stationary transport quantities without feedback. The jump- and no-jump operators in \eq{master_tunnel} are two-by-two matrices, and we parametrized the bare tunnel rates $\Gamma_{L/R}$ between dot and left/right electron reservoirs by the asymmetry parameter $-1\le a\le 1$ \cite{Gustavsonetal06} via $\Gamma_R = \Gamma_L\frac{1-a}{1+a}$ (transport from left to right is assumed). Correspondingly, we introduced an asymmetry parameter $-1\le b\le 1$ as $g_R = g_L\frac{1-b}{1+b}$, where $g_\alpha$ is the dimensionless feedback coupling in the feedback modulated tunnel rates 
$
\Gamma_\alpha(n,t)\equiv \Gamma_\alpha \times f(1+g_\alpha(I_{0}t-n)$, $  \alpha=L/R$.
\begin{figure}[t]
\includegraphics[width=\columnwidth]{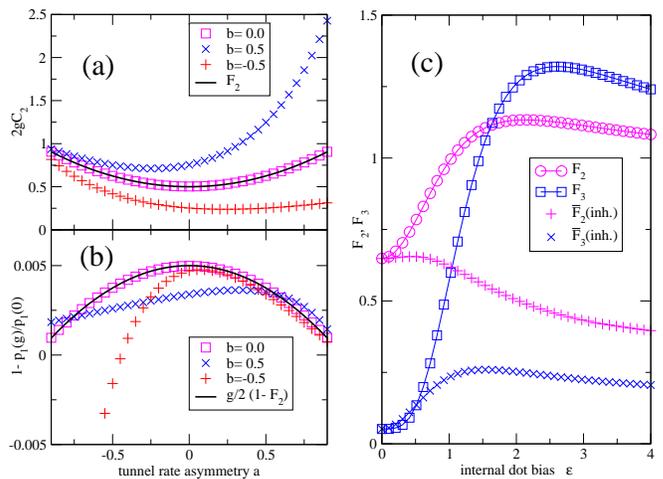}
\caption[]{\label{figure2}
Single quantum dot: (a) second cumulant $C_2$ of frozen FCS with right/left tunnel rates $\Gamma_R = \Gamma_L\frac{1-a}{1+a}$ and feedback couplings $g_R = g_L\frac{1-b}{1+b}$,  $g\equiv g_L=0.02$; (b) relative change of single dot occupation $p_1$;  double quantum dot (c) with internal coupling $T_c=1$, tunnel rates $\Gamma_L=10$, $\Gamma_R=1$, feedback $g=0.01$: 
Fano factor and skewness for homogeneous ($F_2$, $F_3$) and inhomogeneous feedback ($\bar{F}_{2},\bar{F}_{3}$).
}
\end{figure} 

In Fig. (\ref{figure2}a) , we tested  \eq{fano_predict} as a function of $a$: For homogeneous coupling $b=0$, the numerical large-time feedback results precisely match the $g=0$ Fano factor $F_2 = \frac{1}{2}(1+a^2)$. For inhomogeneus feedback $b\ne 0$, this is no longer the case, as could be expected:  the feedback and non-feedback parts of the super-operators in \eq{master_tunnel}
 are not proportial to each other, and the corresponding operators  in the Fourier transformed equation can no longer be replaced by  a single eigenvalue $\lambda_0(\chi)$. The asymmetry parameter $b$ can now be used to fine-tune the 
frozen FCS: for example, negative (positive) $b$ yields an overall stronger (weaker) feedback and thus leads to sharper (broader) feedback distributions. 
We found a similar scenario for transport through a double quantum dot, where we reconstructed  the known results for the $g=0$ Fano factor $F_2$ \cite{Elattari02} and the skewness $F_3$ from the  numerically obtained frozen homogeneous feedback cumulants  via \eq{relation}, cf. Fig. (\ref{figure2}c). In contrast, inhomogeneous feedback that modulates left, right and internal tunnel rates (but not the internal bias $\varepsilon$) leads to corresponding expressions $\bar{F}_{2/3}$ with deviations from  ${F}_{2/3}$ that provide a measure of feedback inhomogeniety.

In contrast to the FCS, the internal state of the system (i.e., the density operator  $\sum_{n=0}^\infty\rho^{(n)}(t)$) is barely changed by the presence of a feedback loop. This can be seen in  Fig. (\ref{figure2}b) where we show the relative change of the single-dot occupation $p_1$ in the stationary limit. For homogeneous feedback $b=0$, this change can  be calculated analytically by inserting the form $\rho(\chi,t)=e^{i\chi I_{0}t}r(\chi)$ into \eq{PDE} and expanding in the feedback coupling $g$, from which equations for the components of $r'(0)$ follow with the result
\be
1- \frac{p_1(g)}{p_1(g=0)}= \frac{g}{2}\left(1-F_2\right),
\ee
thus directly relating a (no-feedback) transport quantity (the Fano factor $F_2$) with a (feedback) occupation probability.
For inhomogeneous feedback, the small modulation of $p_1$ again strongly depends on the sign of $b$.

The feedback mechanism discussed here leads to non-decaying FCS distributions.  A potential application could be the highly accurate transfer of single  electrons at minimal errors. An obvious strong competitor of such a feedback-controlled device are single-electron pumps (or turnstyles, as we are interested in large voltage bias here) that are already used for metrological purposes \cite{pump_metro}. 
We made this comparison more quantitative by modelling a single-electron turnstyle as a single level dot with tunnel rates  rates $(\alpha =L,R)$,
$
\Gamma_\alpha(t) = \gamma_\alpha T\sum_{j=1}^\infty \delta(t-t_{\alpha,j}),
$
with an alternating opening of left and right barrier at times $t_{\alpha,j}$ such that the time elapsed after 
$k$ rounds (left-right-left) is $k \times T$ with $T >0$ the period of the turnstyle. 
We compare the turnstyle with the corresponding feedback-controlled single-level dot, 
under the condition that both configurations transfer the same average charge per time from left to right. At large times, this equivalence leads to 
$
\frac{\Gamma T}{2}= \tanh\left(\frac{\gamma T}{2}\right)
$,
where we assumed symmetrical tunnel rates $\gamma_\alpha = \gamma$ and $\Gamma=\Gamma_\alpha$ is the bare rate in \eq{rates}. 
At large times, the turnstyle with its ever broadening FCS thus eventually becomes inferior to the (ideal) feedback-controlled dot. We define this transition by the time $t^*$ where the second FCS cumulants $C_2$ of both devices coincide. This condition yields
\be\label{tstar}
\Gamma t^* = \frac{1}{g} \cosh^2\left(\frac{\gamma T}{2}\right) ,
\ee
where we used \eq{fano_predict} and $F_2=\frac{1}{2}$ for the feedback $C_2^{\rm fb}$, and the result $C_2^{\rm pump}(t) =  \frac{t}{T} N$, $N\equiv { \tanh\left(\frac{\gamma T}{2}\right)}/(2 \cosh^2\left(\frac{\gamma T}{2}\right))$ that is obtained by considering the change of the Fourier-transformed density operator $\rho(\chi)$ of the turnstyle during one cycle.  Interpreting $N$ as the number of electrons transferred through the turnstyle before the first counting error occurs, \eq{tstar} determines the number of electrons $n^*\equiv \Gamma t^*$ transferred through the feedback device before it becomes superior to the turnstyle, which for $\gamma T \gg 1$ thus happens after  $O(1/g)$ turnstyle counting errors. 

Finally, we address the question of finite delay times that are unavoidable in realistic feedback loops. In general, delay effects in transport will lead to non-Markovian feedback Master equations that generalize the r.h.s. of \eq{master_tunnel} to integrals and sums over kernels at earlier times $t'<t$ and smaller particle numbers $n'<n$, plus additional inhomogeneous terms \cite{BraggioFlindt}. We estimated time-delay effects only, using 
a delay function $\Delta(t') = e^{-t'/\tau}/\tau$  with delay time $\tau>0$  in the feedback Master equation $\dot{\rho}^{(n)}(t) =-\Gamma \int_0^t dt' \Delta(t') [f(q_n(t-t')) \rho^{(n)}(t-t')- (n\to n-1)]$ of the tunnel junction  model.  Laplace-transforming the equations for the first two moments, we obtained the first two cumulants in the long time limit, $C_1(t\to \infty)=\Gamma t$ (which co-incides with the result for zero delay $\tau\to 0$), and
$C_2(t\to\infty)=(1/2g) \times (1 - 2  \Gamma  \tau )$.
This indicates that  delays that are short on the time-scale of the inverse tunneling rate ($\Gamma  \tau\ll 1$) do not re-install a linear increase of the cumulants in time $t$,  but only weakly modify the result \eq{cumulantstunnel} for the  frozen-cumulants.

To conclude, our results demonstrate that the information contained in the full counting statistics can be frozen in 
by a feedback loop, which can lead to a strong suppression of fluctuations on long time scales. We expect that these 
predictions can be tested in quantum transport experiments in the near future.

Discussions with W. Belzig, C. Emary, C. Flindt, G. Kie{\ss}lich and G. Schaller are acknowledged. This work was supported by DFG grant 1528/5-2.


\end{document}